# Pros and cons of gaussian filters versus step filters for light pollution monitoring

Alejandro Sánchez de Miguel[1,2,3]

[1]*Environment and Sustainability Institute, University of Exeter, Penryn, Cornwall TR10 9FE, U.K.*
[2]*Depto. de Física de la Tierra y Astrofísica, Fac. CC. Físicas, Universidad Complutense de Madrid, Plaza de las Ciencias 1, E-28040, Spain*
[3]*Instituto de Física de Partículas y del Cosmos, IPARCOS, Madrid, Spain*

*Corresponding Author: A. Sánchez de Miguel (alejasan@ucm.es)

**Abstract**

There is debate about which indicators should currently be used to monitor levels of artificial light pollution. To be most valuable, methods need to be sensitive to variation in the spectral composition of light emissions (which are changing rapidly, particularly through increasing use of light-emitting diode [LED] lamps), to be readily available, to be capable of being used on a large spatial scale and of being deployed rapidly. Two sets of photometric systems are the most spread in the world currently, the RGB colors from DSLR cameras that are based on typical gaussian filters and RGB step filters. The first set of filters are optimum for human perception and calculation of most of the most popular environmental impacts although, some of these environmental impacts are better characterized by the step filters.

1. **Introduction**

Light pollution is a growing problem with multiple implications. Measuring light pollution started because of practical reasons by professional astronomers in the 70s. But it was not until the 2000s when the SQM photometers started that the massive data collection started. One of the problems that was detected was that the SQM were unable to detect the LED transition producing potentially misleading readings. Meantime, some light pollution experts started to observe that the current metrics on light pollution were inadequate for measuring the environmental and health impacts of the light pollution. Some new metrics were developed in order to improve the situation although currently none of the current metrics has been established as standard. On the other hand, the technical possibilities to develop devices able to measure these metrics without a large investment are limited and photometry is the most promising compared to spectroscopic techniques because cameras and photometers are much cheaper and easy to operate. Also, digital cameras are already deployed all over the world on mobile phones, DSLR cameras, compact cameras, etc. Photometers are another potential solution, as any kind of filter can be added to these devices and there are several networks of photometers already deployed, such as the mentioned SQM photometers, TESS, …. As the case is urgent so we should be able to trace as soon as possible if the new LEDs are effectively reducing or increasing the light pollution, we need to use sets of filters already in production rather than developing new sets of filters especially developed for this purpose. Two kinds of filters exist on the market, RGB gaussian filters and RGB step filters. The first ones are common on cameras and on Johnson astronomical bands, the seconds are common for astrophotography and modern astrophysics studies.

2. **Indicators**

A regulation, to be easy to be implemented, must be well defined and easily enforced. But, as important as it has been easy to be implemented, must be a regulation that comes from common sense and not purely arbitrary. With that idea in mind, we should think that our set of filters that might be used for regulation enforcement, has to be practical to measure also other environmental indicators and not just a nonspecific criterion. That is why in this paper we explore how these filters can be useful for different indicators.

Regulatory indicators: Current regulations are currently based mainly on the photopic human vision and human perception. Illumination levels are measured using photopic curves. The CIE 1931 is used to calculate the CCT that appears on several regulations like the French law, and the Slovenia law. Some observatories have laws that limit the amount of light above a certain wavelength, in this case we are going to consider the 500 nm break. One case of this one is the G index, that is a renormalization of this index and was mentioned in the draft of the Andalucian regulation and European recommendations but has not been implemented yet.

Health indicators: the CIE developed a standard based on the work of Lucas 2014 to study the inputs on the human physiology, in particular the circadian rhythms. There is discussion about if there is an additional input sensitive to UV light. Before this standard, Aube et. al. developed the MSI indicator that has been correlated with higher risk of cancer risk. Also, glare is another important factor that must be taken into consideration.

Environmental indicators: Several spectral factors have been considered for example in Thomas et. al. and Longcore et. al. for several animal species. Also, Aubé et. al. did consider the Induced Photosynthetic Index that might be relevant for plants.

Heritage: The human vision of the stars is of great importance for several cultures and is recognized as all humanity heritage. The human vision of the stars can be estimated by the scotopic vision or the SLI by Aubé et. al.

As filters we have chosen some of the most popular of each kind. As an example of gaussian filters we have used the Nikon D3s and as step filters we have used the RGB filters of Baader planetarium.

https://www.teleskop-express.de/shop/product_info.php/language/en/info/p568_Baader-CCD-Filter-Gruen---1-25----Interferenzfilter.html

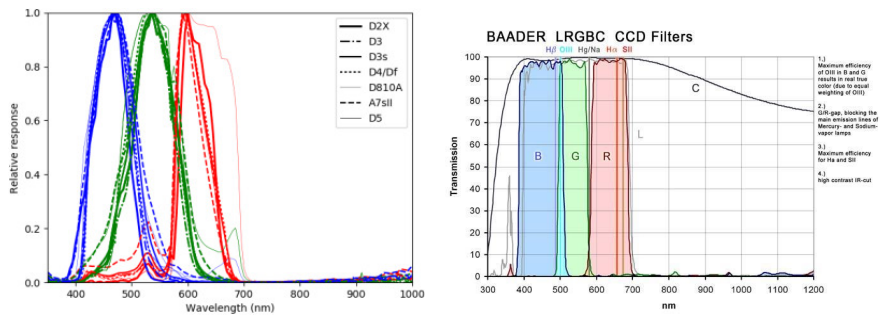

Fig 1. Left gaussian filters, right step filters. Left gaussian filters from several manufacturers although focused mainly on the Nikon ones. right panel an example of step fitters from the brand Baader planetarium.

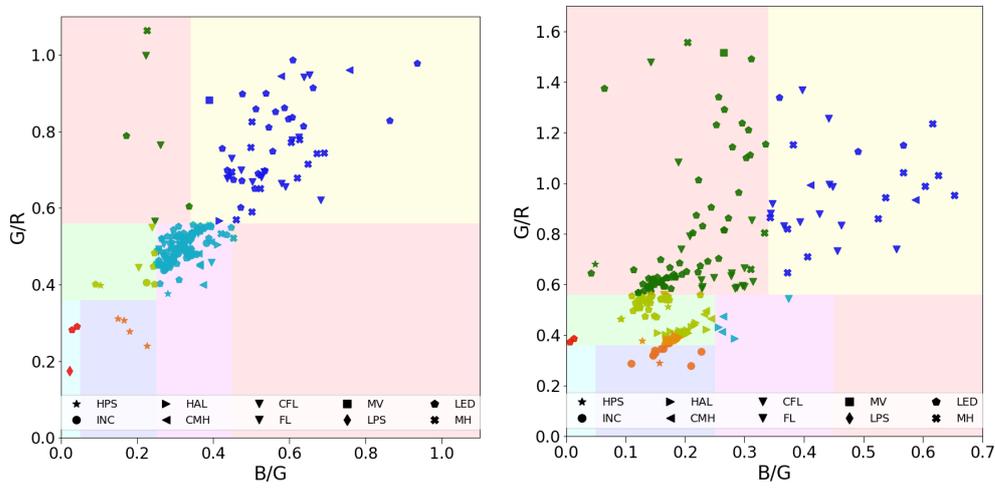

Fig 2. Left gaussian filters, right step filters. Both plots show that there is overlap between technologies and is impossible to separate completely the different technologies. Although. is clear that if the white sources like FL and MH where very mixed on the left panel and the Sodium plights where clearly separated, on the right panel is the LEDs and the MH which ones are separated, and the sodium lights are mixed with LEDs and incandescent and Halogen.

Environmental indicators.

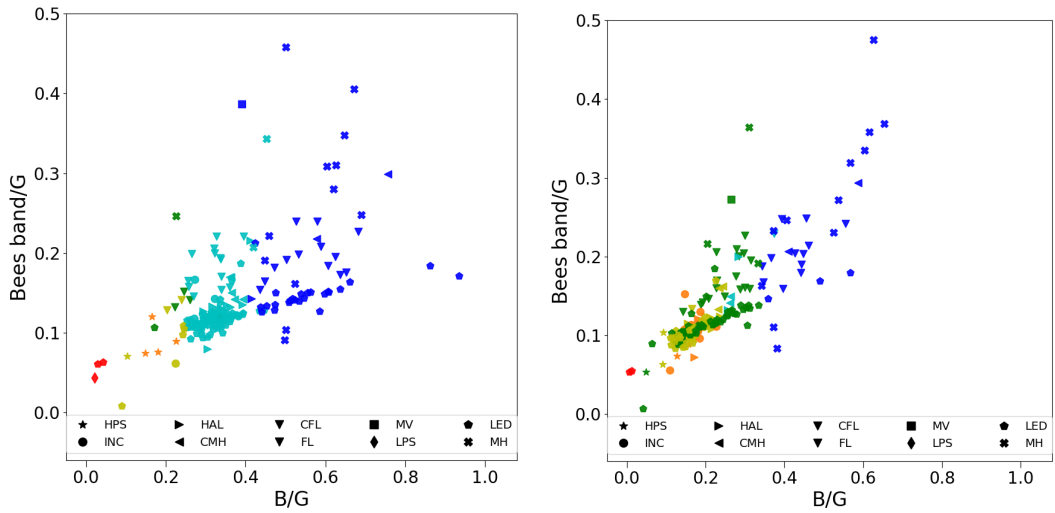

Fig 3. Left gaussian filters, right step filters. Here can be seen the impact of the Bees attraction band compared to gaussian and step filters. both present clear correlation, although the right plot is less scattered on the bluish part, is more confused on the reddish part.

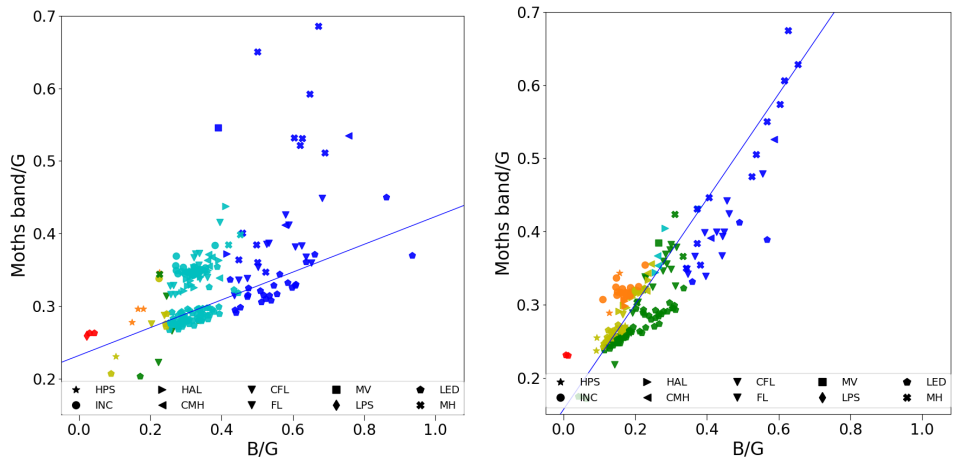

Fig 4. Left gaussian filters, right step filters. Here can be seen the impact of the Moths attraction band compared to gaussian and step filters. both present clear correlation, although the right plot is less scattered on the bluish part, is more confused on the reddish part.

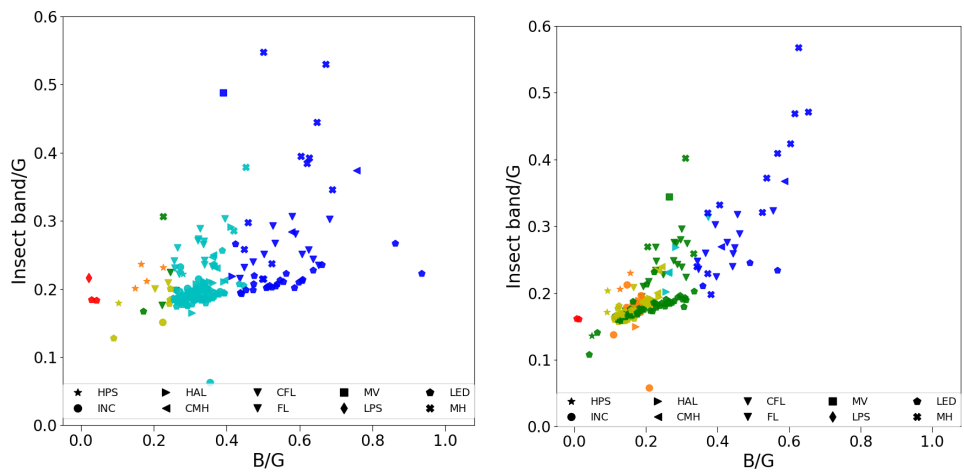

Fig 5. Left gaussian filters, right step filters. Here can be seen the impact of the Insect attraction band compared to gaussian and step filters. The gaussian filter do not present clear correlation but the step filter does.,

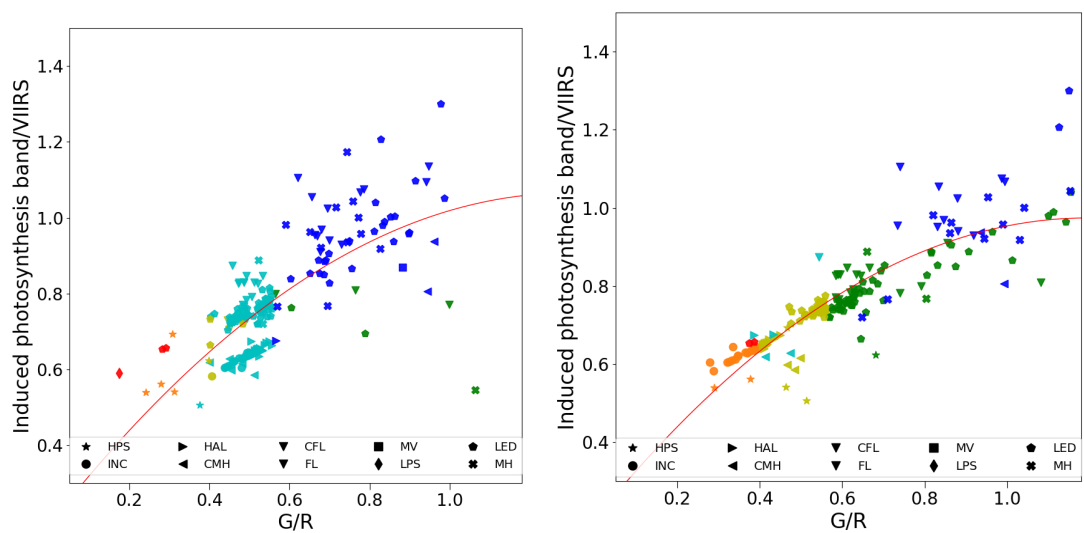

Regulations

Fig 6. Left gaussian filters, right step filters. Here can be seen the impact of the Induced Photosynthesis Index compared with VIIRS data. Both present correlation although the step is much less scattered.

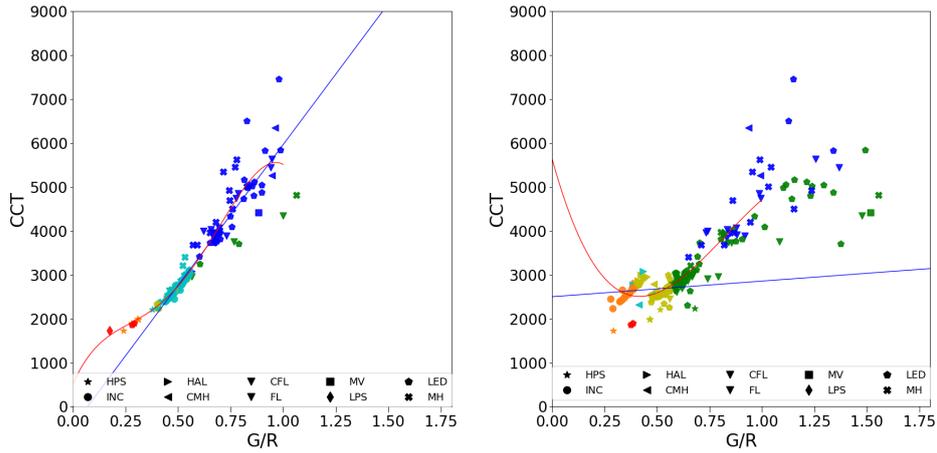

Fig 7. Left gaussian filters, right step filters. Here can be seen the impact of the CCT. Left is very well constrained, but right is very diffuse.

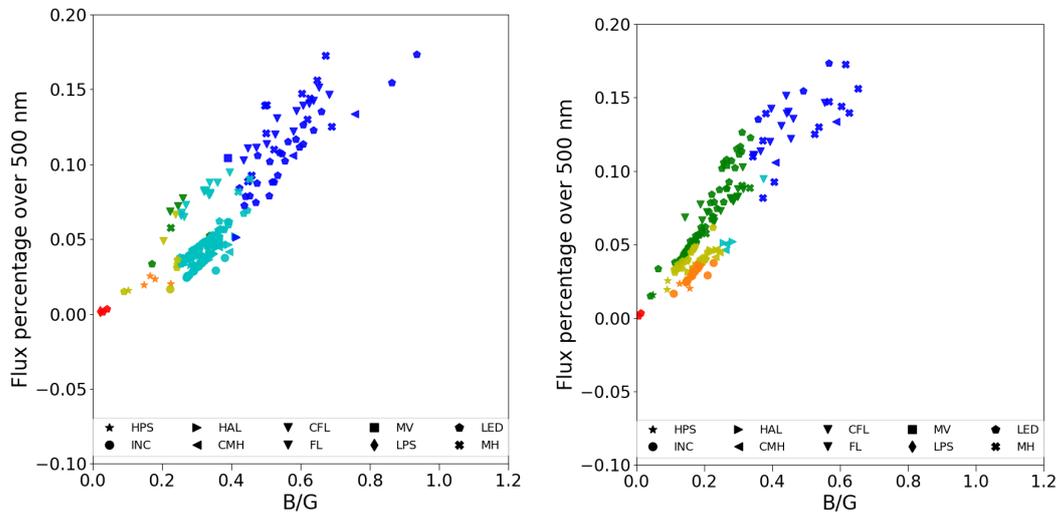

Fig 8. Left gaussian filters, right step filters. Here can be seen the impact of the blue percentage (equivalent to the G index), both present correlation but right is more precise except there is more confusion on the reddish area.

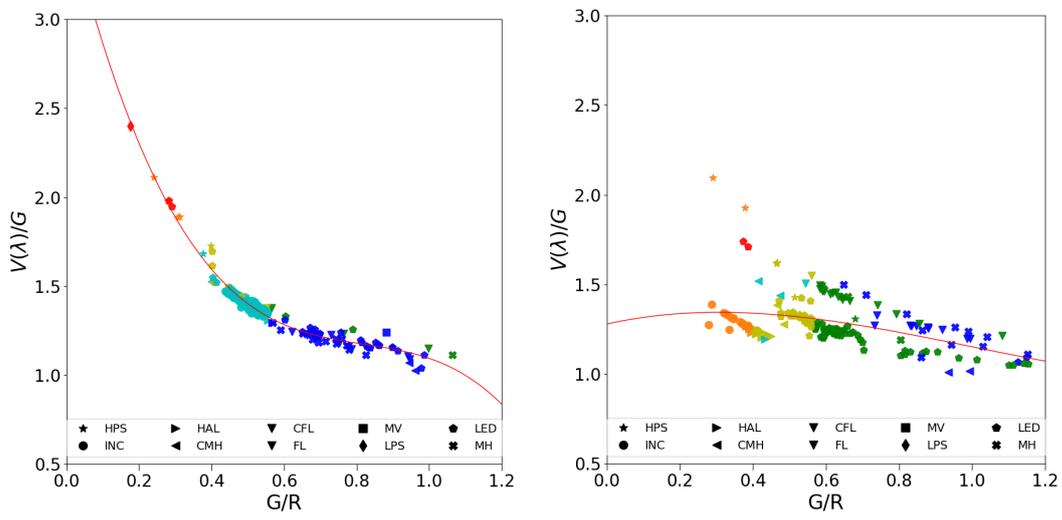

Fig 9. Left gaussian filters, right step filters. Photopic.

Human health

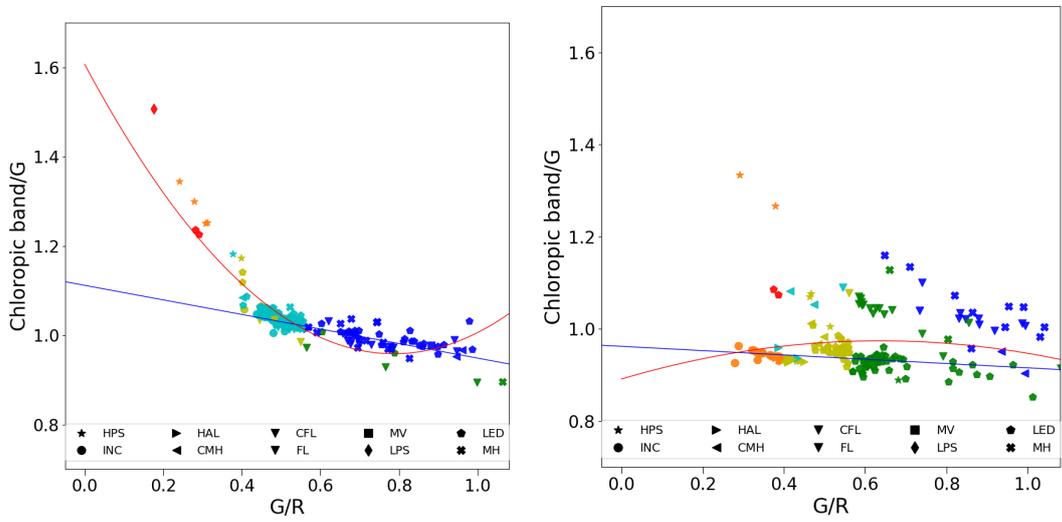

Fig 10. Left gaussian filters, right step filters.

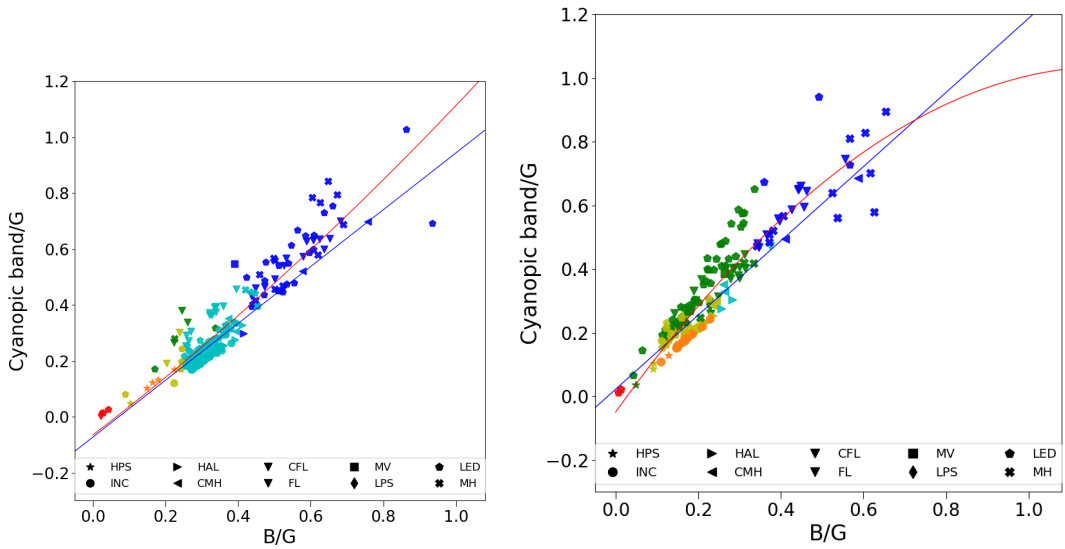

Fig 11. Left gaussian filters, right step filters.

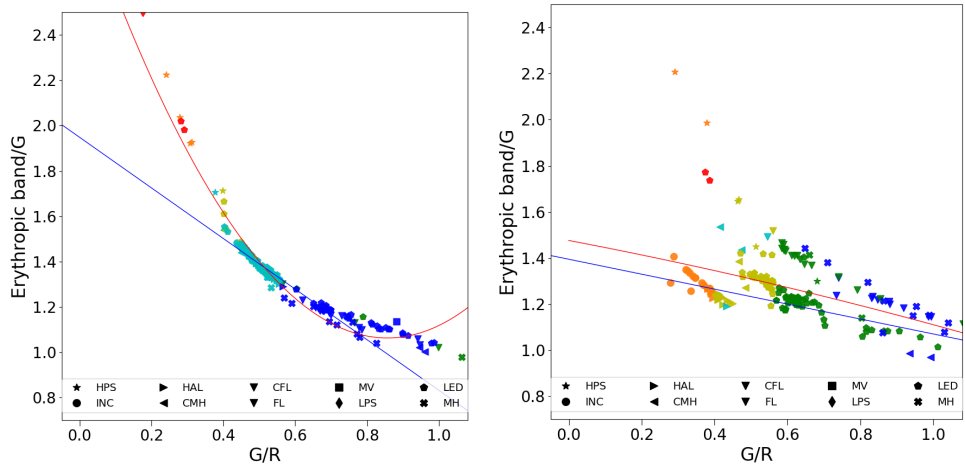

Fig 12. Left gaussian filters, right step filters.

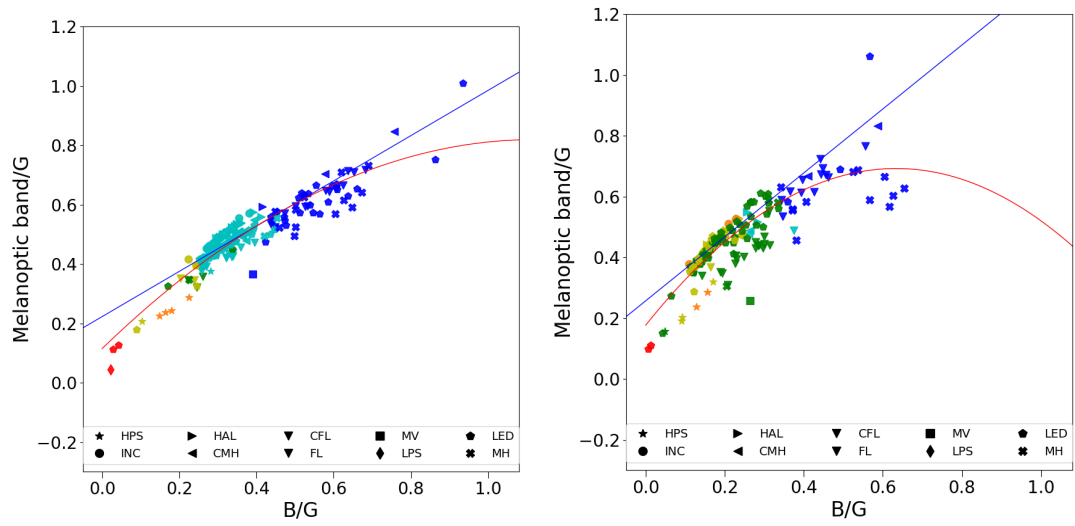

Fig 13. Left gaussian filters, right step filters.

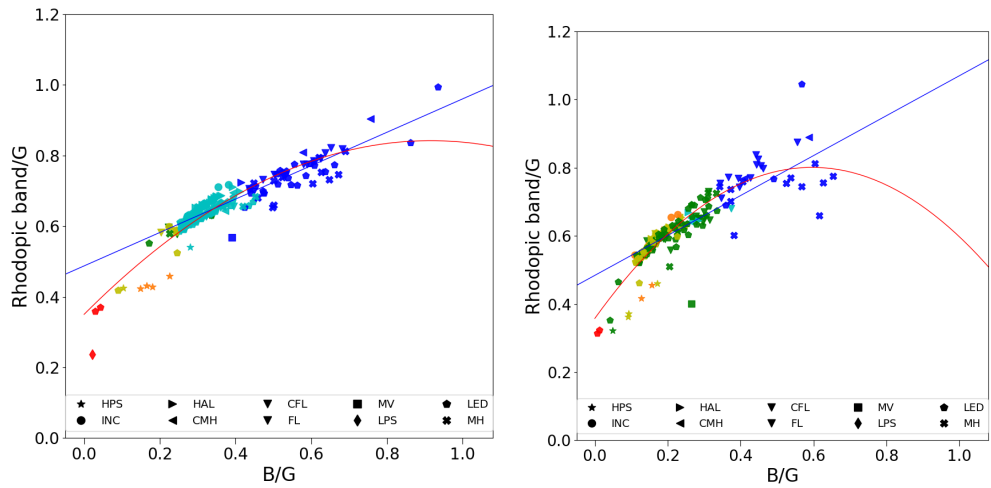

Fig 14. Left gaussian filters, right step filters.
Heritage

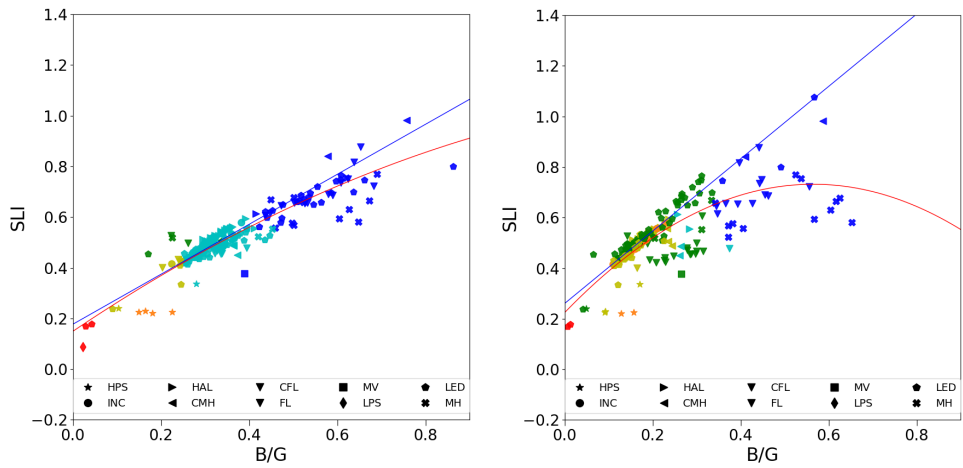

Fig 15. Left gaussian filters, right step filters. SLI

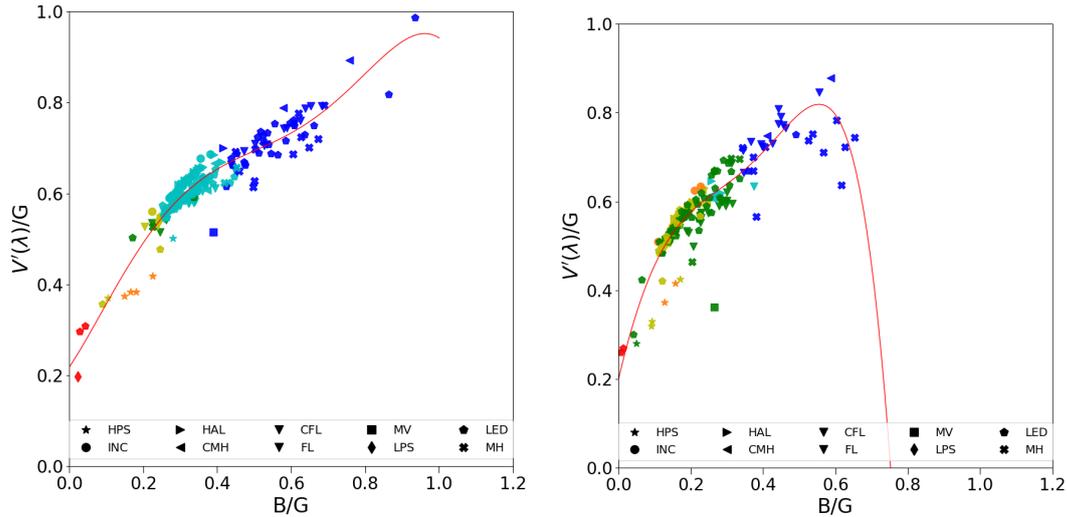

Fig 16. Left gaussian filters, right step filters. Scotopic

**1.**

## 2. Methods

### 2.1. Data

For this work we have used the same data from Sánchez de Miguel et. al. 2019 and Sánchez de Miguel et. al. 2019. The mathematical methods are also described in Sánchez de Miguel et. al. 2019. and Sánchez de Miguel et. al 2020.

## 3. Results

### 3.1. Descriptive statistics

The Step filters perform better on the estimation of the G index, Insect, bees and moths. Perform slightly worse on the melatonin or MSI index. And the gaussian filters clearly are better for CCT, photopic, scotopic and all human physiology.

|        | Step filters |      |      |       |      |       |       |      |      |      |      |      |      |
|--------|------|------|------|-------|------|-------|-------|------|------|------|------|------|------|
|        | BG   | GR   | RG   | B/RGB | B/GR | G/RGB | R/RGB | RGB1 | RGB2 | RGB3 | RGB4 | RGB5 | RGB6 |
| B500   | 0,86 | 0,76 | 0,68 | 0,97  | 0,97 | 0,56  | 0,82  | 0,95 | 0,95 | 0,97 | 0,97 | 0,97 | 0,94 |
| MSI    | 0,86 | 0,63 | 0,56 | 0,92  | 0,92 | 0,41  | 0,71  | 0,91 | 0,88 | 0,92 | 0,92 | 0,92 | 0,91 |
| SLI    | 0,63 | 0,48 | 0,37 | 0,64  | 0,63 | 0,31  | 0,46  | 0,60 | 0,58 | 0,60 | 0,60 | 0,60 | 0,58 |
| Vl/G   | 0,17 | 0,37 | 0,41 | 0,24  | 0,23 | 0,35  | 0,37  | 0,37 | 0,27 | 0,33 | 0,33 | 0,33 | 0,18 |
| SC/G   | 0,78 | 0,35 | 0,18 | 0,67  | 0,65 | 0,18  | 0,32  | 0,73 | 0,72 | 0,70 | 0,70 | 0,70 | 0,61 |
| CCT    | 0,61 | 0,82 | 0,78 | 0,79  | 0,79 | 0,61  | 0,87  | 0,88 | 0,77 | 0,85 | 0,85 | 0,85 | 0,78 |
| rhoG   | 0,80 | 0,35 | 0,19 | 0,69  | 0,67 | 0,18  | 0,34  | 0,74 | 0,74 | 0,71 | 0,71 | 0,71 | 0,62 |
| melG   | 0,74 | 0,36 | 0,18 | 0,63  | 0,62 | 0,19  | 0,31  | 0,69 | 0,69 | 0,68 | 0,68 | 0,68 | 0,60 |
| CyaG   | 0,88 | 0,68 | 0,57 | 0,94  | 0,94 | 0,47  | 0,73  | 0,92 | 0,93 | 0,93 | 0,93 | 0,93 | 0,91 |
| Chl/G  | 0,10 | 0,18 | 0,16 | 0,04  | 0,03 | 0,21  | 0,13  | 0,31 | 0,19 | 0,25 | 0,25 | 0,25 | 0,02 |
| moths/G| 0,83 | 0,20 | 0,11 | 0,68  | 0,68 | 0,06  | 0,25  | 0,89 | 0,91 | 0,88 | 0,88 | 0,88 | 0,62 |
| Insect/G| 0,73| 0,19 | 0,16 | 0,66  | 0,67 | 0,08  | 0,28  | 0,70 | 0,70 | 0,67 | 0,67 | 0,67 | 0,61 |
| psas/G | 0,06 | 0,85 | 0,81 | 0,17  | 0,17 | 0,92  | 0,73  | 0,69 | 0,97 | 0,87 | 0,87 | 0,87 | 0,09 |

Table 1. R2 of bidimensional and tridimensional fits to different indicators using RGB step filters. Red is between 0.7 and 0.8, Orange between 0.8 to 0.9 and Green 0.9 and 1.0. RGBX are different combinations of RGB. B500, blue light above 500 nm; MSI, Melatonin Suppression Index; SLI, Vl/G Photopic ratio Green band; SC/G, Scotopic ratio Green band; CCT, Color Correlated Temperature; rhoG, rhodopic ratio Green band;melG, Melanopic ratio green band; ChlG, Chloropic Green ratio; Moth band

|        | Gaussian filters |      |      |       |      |       |       |      |      |      |      |      |      |
|--------|------|------|------|-------|------|-------|-------|------|------|------|------|------|------|
|        | BG   | GR   | RG   | B/RGB | B/GR | G/RGB | R/RGB | RGB1 | RGB2 | RGB3 | RGB4 | RGB5 | RGB6 |
| B500   | 0,83 | 0,72 | 0,67 | 0,85  | 0,85 | 0,52  | 0,78  | 0,83 | 0,80 | 0,83 | 0,83 | 0,83 | 0,83 |
| MSI    | 0,89 | 0,74 | 0,71 | 0,91  | 0,91 | 0,55  | 0,83  | 0,91 | 0,89 | 0,91 | 0,91 | 0,91 | 0,91 |
| SLI    | 0,87 | 0,74 | 0,74 | 0,92  | 0,91 | 0,64  | 0,85  | 0,88 | 0,90 | 0,88 | 0,88 | 0,88 | 0,85 |
| Vl/G   | 0,72 | 0,97 | 0,97 | 0,86  | 0,84 | 0,92  | 0,98  | 0,76 | 0,97 | 0,87 | 0,87 | 0,87 | 0,61 |
| SC/G   | 0,90 | 0,64 | 0,64 | 0,90  | 0,88 | 0,58  | 0,75  | 0,82 | 0,85 | 0,77 | 0,77 | 0,77 | 0,73 |
| CCT    | 0,82 | 0,91 | 0,90 | 0,91  | 0,91 | 0,69  | 0,98  | 0,95 | 0,82 | 0,92 | 0,92 | 0,92 | 0,92 |
| rhoG   | 0,90 | 0,67 | 0,67 | 0,91  | 0,89 | 0,61  | 0,78  | 0,81 | 0,87 | 0,77 | 0,77 | 0,77 | 0,74 |
| melG   | 0,90 | 0,57 | 0,56 | 0,88  | 0,87 | 0,52  | 0,70  | 0,85 | 0,86 | 0,79 | 0,79 | 0,79 | 0,76 |
| CyaG   | 0,89 | 0,72 | 0,66 | 0,89  | 0,89 | 0,51  | 0,78  | 0,88 | 0,87 | 0,88 | 0,88 | 0,88 | 0,87 |
| Chl/G  | 0,50 | 0,92 | 0,91 | 0,67  | 0,63 | 0,93  | 0,88  | 0,51 | 0,91 | 0,75 | 0,75 | 0,75 | 0,29 |
| moths/G| 0,39 | 0,19 | 0,16 | 0,34  | 0,35 | 0,10  | 0,24  | 0,40 | 0,40 | 0,38 | 0,38 | 0,38 | 0,35 |
| Insect/G| 0,29| 0,21 | 0,18 | 0,28  | 0,28 | 0,11  | 0,24  | 0,25 | 0,25 | 0,25 | 0,25 | 0,25 | 0,25 |
| psas/G | 0,12 | 0,27 | 0,27 | 0,16  | 0,15 | 0,31  | 0,24  | 0,29 | 0,14 | 0,24 | 0,24 | 0,24 | 0,13 |

Table 2. R2 of bidimensional and tridimensional fits to different indicators using RGB gaussian filters. Red is between 0.7 and 0.8, Orange between 0.8 to 0.9 and Green 0.9 and 1.0. RGBX are different combinations of RGB.

|    | B500 | MSI  | SLI  | Vl/G | SC/G | CCT  | rhoG | melG | CyaG | Chl/G | moths/G | Insect/G | psas/G |
|----|------|------|------|------|------|------|------|------|------|-------|---------|----------|--------|
| S  | 0,97 | 0,92 | 0,63 | 0,41 | 0,78 | 0,88 | 0,8  | 0,74 | 0,94 | 0,31  | 0,91    | 0,73     | 0,97   |
| G  | 0,85 | 0,91 | 0,92 | 0,97 | 0,9  | 0,98 | 0,91 | 0,9  | 0,89 | 0,93  | 0,39    | 0,29     | 0,31   |
| Op | S    | =    | G    | G    | G    | G    | G    | G    | S    | G     | S       | S        | S      |

Table 3. Comparison between best fit of Step filters and gaussian filters for different indicators.

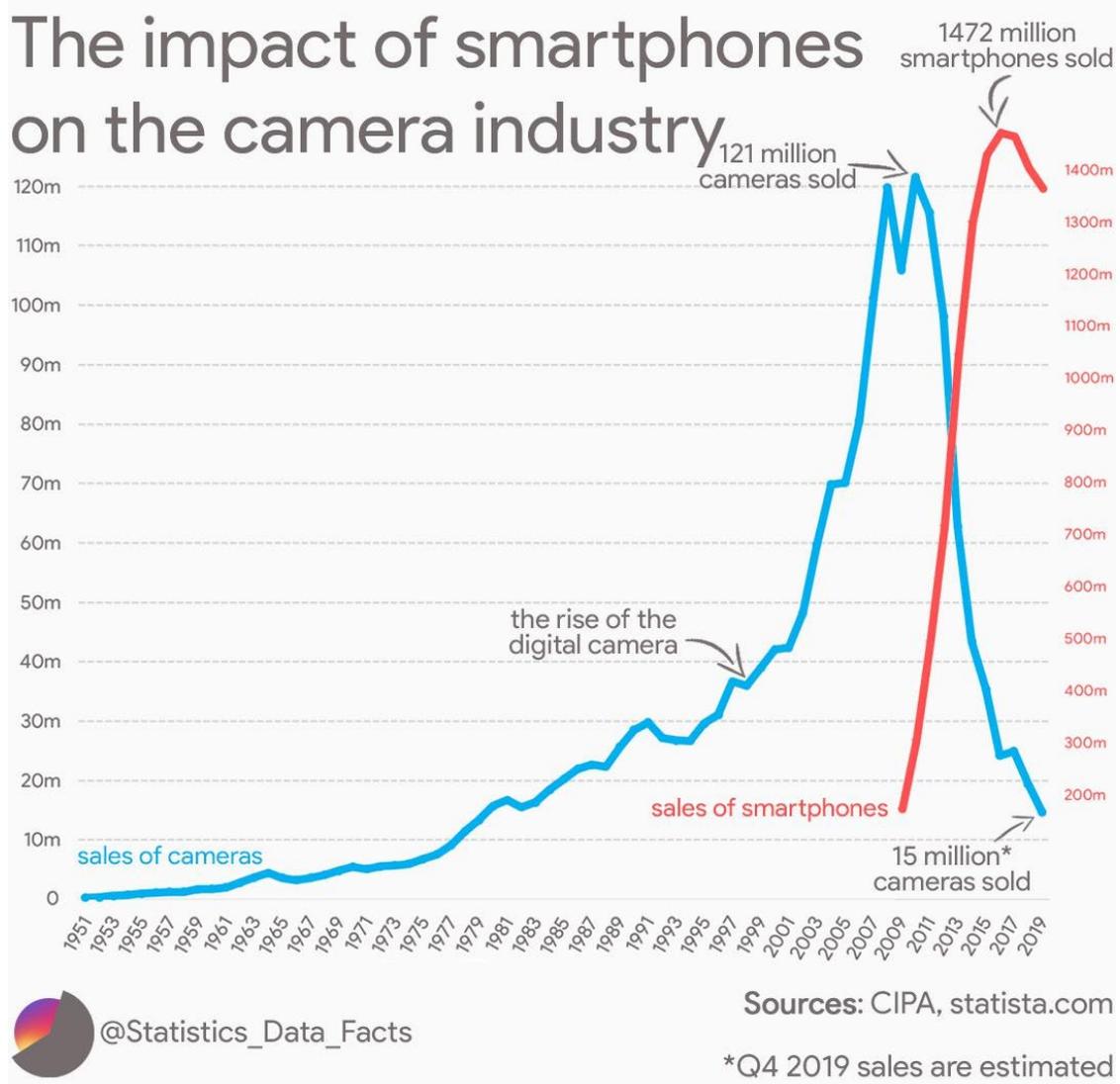

Fig 17. Popularity of different sensors.

(https://petapixel.com/2017/03/03/latest-camera-sales-chart-reveals-death-compact-camera/)

4. **Discussion**

Both sets of filters have their weaknesses and strengths. The Step filters are easier to define but more difficult to deploy and are bad filters for most of the current regulations although for some like the G index are perfect. Again, they are very bad for human health and human vision estimations, with the exception of the neuropsin, although the use of this pigment is under discussion. Also, they are more difficult to deploy as they need to use cameras or photometers specifically designed for that purpose. On the other hand, the gaussian filters perform very well on the human vision metrics, health metrics with exception of the neuropsin. The performance on other environmental indicators is clearly worse than the step filters although they still can give some information. Only for the G index and the mentioned neuropsine, these filters are very badly performing. Their main advantage is that they are present on all DSLR cameras and mobile phones.

If the current situation makes that the speed of deployment is the main factor, clearly the gaussian filters should be used as standard. Similar arguments can be made if we consider backwards compatibility with current laws. That does not mean that the step filters are not useful, they have their place as ideal tools to estimate some environmental indicators but are not suitable now for a massive change. If they are chosen as standard, it would force duplicate resources for many professionals and will produce that their implementation is slower. On fig 17 can be seen the potentially available sensors with RGB gaussian filters. The number of devices with step filters can be considered neglectable, as those kinds of filters are only used on astrophotography and they are not built on the self of the devices, they are used as external filters for panchromatic cameras.

So, the step filters are the exception on the current daily life sensors and the gaussian filters are the standard. The astronomers did use gaussian filters in the past as the Johnshon cousins filters, but for many applications like hyperspectral observations, customized filters are advised. Should the regular people fit on the astronomical standards even if they perform worse on some of the most popular indicators? Clearly if we want to control light pollution worldwide, the answer is no. Although, astronomical techniques like the explained in Sánchez de Miguel et. al. 2019, 2021 and Galadí-Enriquez 2019 are essential to design tools for the cheap and fast calibration and control of light pollution, not only for daily live use but also for other scenarios like environmental impact studies.

## 7. Summary and Conclusions

Both sets of filters have their best use. The step filters are useful although their lack of compatibility and less current deployment make them suitable than the gaussian ones. This fact joint with the fact that the abundance of the gaussian filters on the market versus the inexistence of the step filters on the daily life devices make the gaussian filters the best option for a base light pollution control metric.